\documentstyle[aas2pp4,epsf]{article}

\lefthead{W. L. Holzapfel \ea }
\righthead{Non Detection}

\slugcomment{To be submitted to {\em Astrophysical Journal}}

\def\kl{{\rm{k}\lambda}}
\def\ea{et al.\ }
\def\ah{^{\rm h}}
\def\am{^{\rm m}}
\def\as{^{\rm s}}
\def\pr{^{\prime}}
\def\2pr{^{\prime \prime}}
\def\deg{^{\circ}}
\def\greatsim{\mathrel{\raise.3ex\hbox{$>$\kern-.75em\lower1ex\hbox{$\sim$}}}}
\def\lesssim{\mathrel{\raise.3ex\hbox{$<$\kern-.75em\lower1ex\hbox{$\sim$}}}}

\begin{document}

\title{A Search for CMB Decrements Towards Distant Cluster Candidates PC1643+4631 
and VLA1312+4237 at $28.5\,$GHz}

\author{
W.L.~Holzapfel\altaffilmark{1},
J.E.~Carlstrom\altaffilmark{2},
L.~Grego\altaffilmark{3},\\
M.~Joy\altaffilmark{4},
and E.D.~Reese\altaffilmark{2}}


\altaffiltext{1}{Department of Physics, University of California,
Berkeley, CA 94720}
\altaffiltext{2}{Department of Astronomy, University of Chicago, Chicago
IL 60637}
\altaffiltext{3}{Harvard-Smithsonian Center for Astrophysics, Mail Stop 83,
60 Gardeen St., Cambridge MA 02138}
\altaffiltext{4}{Space Science Laboratory, SD50, NASA Marshall Space Flight Center, 
Huntsville AL 35812}
\altaffiltext{5}{Department of Physics, University of Alabama, Huntsville AL
35899}
\authoremail{swlh@cfpa.berkeley.edu}

\begin{abstract}
Recently, \markcite{Jones}Jones \ea (1997) used the Ryle telescope, operating at a 
frequency of $15\,$GHz, to detect a flux decrement in the direction of the 
quasar pair PC1643+461A,B.
They interpreted this signal as the Sunyaev-Zel'dovich effect (SZE) produced by a 
distant cluster of galaxies.
In the course of an effort to measure CMB anisotropies using the VLA at $8.4\,$GHz,
\markcite{Richards}Richards \ea (1997) detected a similar, but smaller, decrement
which we refer to as VLA1312+4237.
They also proposed that this signal might be explained as the SZE
signal of a distant galaxy cluster.
We report observations in the direction of these claimed sources with the 
Berkeley Illinois Maryland Association (BIMA) interferometer operating at 
$28.5\,$GHz.
We find no evidence for SZE emission in the direction of either of the 
claimed sources.    
In the case of PC1643+4631, the BIMA data are inconsistent with the cluster 
emission model proposed by \markcite{Jones}Jones \ea (1997) at 
greater than $99.99\%$ confidence.
Together with published x-ray and optical searches, these results make a 
compelling case against the existence of a massive cluster in the 
direction of PC1643+4631.
Because of the different scales to which the VLA and BIMA instruments are sensitive, 
the BIMA observations are not as constraining for the VLA1312+4237 source.
The BIMA data are inconsistent with the cluster model proposed by 
\markcite{Richards}Richards \ea (1997) at $\sim 80\%$ confidence.
\end{abstract} 

\keywords{cosmology: observation --- cosmic microwave background ---
galaxies: clusters --- techniques: interferometric}

\section{Introduction}
The Sunyaev-Zel'dovich effect (SZE) is the scattering of Cosmic Microwave 
Background (CMB) photons by the hot plasma bound to massive clusters of galaxies.
This scattering results in a spectral distortion of the CMB, observable at 
wavelengths from the radio to the sub-mm.
At frequencies lower than $\sim 217\,$GHz, the SZE is observed as a 
decrement in the temperature of the CMB towards massive clusters of galaxies. 
Because the SZE surface brightness of distant clusters is independent of
redshift, radio surveys have the potential to be a particularly powerful 
method of locating distant galaxy clusters.  
In low density universes, clusters are presently not evolving rapidly and 
should exist relatively unchanged to high redshift.
The ability of cluster number counts, especially at high redshift, to distinguish 
between different cosmological models has been extensively discussed; 
for example see \markcite{Barbosa}Barbosa \ea (1998).
 
Significant detections of the SZE have been obtained for dozens of 
x-ray and optically selected clusters (see \cite{Birkinshaw} for a review).
Historically, the sensitivity and sky coverage of radio surveys have not
been sufficient to allow the detection of unknown clusters through their SZ
signal.
However, recent advances in detector technology and observing strategy have 
placed the goal of using the SZE to search for distant
unknown clusters within reach.
Recently, two groups working with sensitive interferometers have
claimed the detection of significant small angular scale decrements
in the CMB which are not in the direction of known clusters of 
galaxies.
These detections, if they were due to very distant clusters, 
would be of profound cosmological significance (\cite{Bartlett}).
The lack of supporting evidence for low redshift clusters in the
direction of the claimed decrements, and the difficulty of reconciling 
very high redshift clusters with favored cosmological models, have 
prompted several authors to propose alternative
explanations for the decrements (\cite{Natarajan}; \cite{Dabrowski}). 

We have observed fields containing each of the claimed decrements with
the BIMA interferometer, configured with sensitive $28.5\,$GHz receivers,
in an attempt to confirm the detections.
In this work, we review the previous observations of these
fields and explore the constraints that the BIMA results place on
the explanation of the observed decrements being due to the SZE in distant
clusters.

\section{PC1643+4631 Observations}
\subsection{Ryle Telescope Observations}
The Ryle telescope has been used by \markcite{Jones}Jones \ea (1997) to 
image the surroundings 
of a number of radio quiet quasars at a frequency of $15\,$GHz.
The deepest of these images was towards the $z=3.8$ quasar pair 
PC 1643+4631A,B.
The CLEANed map, containing baselines from $1.25-5.4\,{\rm k}\lambda$, had
a {\it rms} noise of $33\,\mu$Jy in a $33\2pr \times 42\2pr$ beam and was used to
search for point sources.
Three point sources with flux densities of $550\,\mu$Jy, $200\,\mu$Jy,
and $150\,\mu$Jy were detected and CLEANed from the visibility data. 
When only baselines shorter than $1.25\,{\rm k}\lambda$ were included, 
the east-west configuration of the Ryle telescope,
produced a north-south elongated beam $110\2pr \times 170\2pr$.
They find an essentially unresolved decrement in the resulting map of 
$-380 \pm 64\, \mu$Jy centered at 
$\alpha = 16\ah\,43\am\,44.0\as\ ,\, \delta= +46\deg\,30\pr\,20\2pr$ (J2000).
The authors interpret their signal as being due to the SZE in a
massive cluster of galaxies. 
Modeling the emission by a spherical King model,
\begin{equation}
\Delta T=\Delta T_0 \left(1 + {{\theta^2}\over{\theta_c^2}}\right)^{-\frac{3}{2} \beta + \frac{1}{2}}
\label{eqn:sking}
\end{equation}
with $\beta=2/3$, they find a minimum value for the central temperature decrement of 
$\Delta T_0=-560\, \mu$K, with a corresponding angular core radius of $\theta_c= 60\2pr$. 

\subsection{X-ray} 
The ROSAT PSPC is sensitive to x-rays in the energy range $0.1-2.4\,$keV, and
is therefore well suited to the study of high redshift clusters.
One of the last observations of the ROSAT PSPC was used to search
for the x-ray radiation that would be expected from a massive cluster
capable of producing the PC1643+4631 CMB decrement (\cite{Kneissl}). 
The total time of exposure was $\sim 16\,$ks.
Within a $1.5\pr$ circular aperture around the reported position of the
central decrement, there were only 13 counts compared to expected 
background of 15.
This result was used to place a limit on the bolometric flux, 
$f_x < 1.9 \times 10^{-14}\,{\rm erg\,cm}^{-2}\,{\rm s}^{-1}$
at $99.7\%$ confidence.
With these results, \markcite{Kneissl}Kneissl, Sunyaev, \& White (1997) established a 
conservative lower limit on the redshift of any isothermal object 
with temperature between $0.2$ and $10\,$keV 
and capable of producing the observed SZE decrement, of $z > 2.8$.
Open models of the universe with $\Omega_m < 0.25$ predict only one 
cluster on the entire sky with redshift $3<z<4$ and mass $M > 2\times 10^{15} M_{\sun}$.

\subsection{Optical}
The field for the radio observations was originally selected  by
\markcite{Jones}Jones \ea (1997) due to the presence of two 
radio quiet quasars PC1643+4631A,B with redshifts $z=3.79 and 3.83$
separated by $198\2pr$.
It was suggested by \markcite{Saunders}Saunders \ea (1997) that, 
despite the difference in redshift, the two quasars might be gravitationally
lensed images of the same object.
If correct, this system would be the largest separation lensed multiple 
image yet discovered. 
The x-ray results place 
any candidate for producing the SZE signal at too high a redshift to 
produce such lensing and another closer lensing mass of 
$\sim 10^{15}\, M_{\sun}$ would be necessary. 

The field of PC1643+4631 has been extensively searched for an excess 
of faint galaxies
which might be expected to be associated with a distant cluster. 
Using the lack of excess galaxies, \markcite{Saunders}Saunders \ea (1997) 
constrained the redshift of any cluster to be $z>1$.
This limit was revised downward somewhat by \markcite{Cotter1998a}Cotter 
\ea (1998b), who used simulations to show that a cluster at redshift $z=1$ 
would be difficult to locate in the PC1643+4631 field.

Deep UVR imaging is an efficient means to find  $z= 3-4$ galaxies.
Multicolor imaging has been used to place limits on the surface 
density number of Lyman-break galaxies at $z>3$ in the direction of 
PC1643+4631 (\cite{Cotter1998b}; \cite{Haynes}). 
These papers claim a $2 \sigma$ excess of galaxies over the number
expected from the work of 
\markcite{Steidel}Steidel \ea (1998) although, as they point out, 
the number of objects leading to this result is small. 
\markcite{Ferreras}Ferreras \ea (1998) have imaged the region surrounding
PC1643+4631A in the UVR bands.
The corner of their $3.8\pr \times 3.8\pr$ frame includes the position
of the reported decrement.
They find that the Lyman-break galaxies are homogeneously distributed
across the frame, and do not clump near the reported position of the
decrement.
This result is reproduced by \markcite{Haynes}Haynes \ea (1998).
In fact, they find an anticorrelation between the position of Lyman-break 
galaxies and the P1643+4631 decrement, although the small number of galaxies
makes the significance questionable. 
Given the magnitude limits of the observations, the authors 
claim it would be unlikely that galaxies at redshift $z=3.81$
could be detected. 
They go on to propose that the dearth of galaxies in the direction of the PC1643+4631 
decrement is consistent with the lensing signature of a massive
cluster at $z\sim 2$. 
However, the x-ray results place any cluster capable of producing the 
claimed decrement at a redshift of $z>2.8$, at which point it would be
incapable of producing the supposed lensing.
Therefore, we conclude that there is no optical evidence for a 
massive galaxy cluster in the direction of PC1643+4631A,B.

\subsection{BIMA and OVRO Observations}
We observed PC1643+4631 with our cm-wave receivers mounted on 9 of the
$6.1\,$m telescopes of the (Berkeley Illinois Maryland Association) BIMA array.
This system has been used to obtain high signal to noise images 
of the SZE toward more than $15$ clusters, and
for more than 20 clusters if we include observations with the same
receivers mounted on the OVRO millimeter-array (see \cite{Carlstrom99}). 

The primary beam at our observing frequency of $28.5\,$GHz is
$6.6\pr$ FWHM.  
To achieve high brightness sensitivity, seven telescopes were
arranged in a compact non-redundant configuration contained within a triangular
region roughly 18 m on a side.  
To improve discrimination against point sources, the remaining two telescopes 
were placed at stations $47\,$m East and $70\,$m 
North from the center of the compact array.
The primary beam and the range of projected baselines ($0.6\,
{\rm k}\lambda$ to $8.3\,{\rm k}\lambda$) are well matched to the $15\,$GHz 
Ryle telescope observations
(\cite{Jones}). The position of the reported decrement, which we
used for our pointing and phase center, is listed in
Table~\ref{tab:sourcepos}.

\begin{table*}
\begin{center}
\begin{tabular}{llcc}
\multicolumn{4}{c}{BIMA Pointing and Source Positions}\\\tableline\tableline
Object & \multicolumn{1}{c}{Position} & $\alpha\,$(J2000) & $\delta\,$(J2000) \\\tableline
PC1643+4631 & BIMA Pointing& $16\ah\,45\am\,11.278\as$ & $+46^{\circ}\,24\pr\,55.8\2pr$ \\
PC1643+4631 & Ryle Source& $16\ah\,45\am\,11.278\as$ & $+46^{\circ}\,24\pr\,55.8\2pr$ \\\tableline
VLA1312+4237 & BIMA Pointing 	& $13\ah\,12\am\,17.397\as$ & $+42^{\circ}\,38\pr\,05.0\2pr$ \\
VLA1312+4237 & VLA North peak	& $13\ah\,12\am\,17.157\as$ & $+42^{\circ}\,37\pr\,44.5\2pr$ \\
VLA1312+4237 & VLA South peak   & $13\ah\,12\am\,17.122\as$ & $+42^{\circ}\,37\pr\,15.0\2pr$  \\
\end{tabular}
\end{center}
\caption[]{BIMA pointing centers and source positions. 
The lines reading `BIMA Pointing' give the coordinates of the pointing and phase 
center for the BIMA observations. The VLA feature is elongated in declination,
we give the positions of the northern and southern sub-peaks.
}
\label{tab:sourcepos}
\end{table*}

The system temperatures, scaled to above the atmosphere, ranged from $35\,$K
to $55\,$K, depending on atmospheric water vapor content and elevation of
the source.  
The signals were combined in the BIMA 2-bit digital correlator configured for 8
contiguous $100\,$MHz sections each with 32 complex channels. 
After bandpass corrections were made and end channels dropped, channels from
each correlator section were averaged to produce eight $100\,$MHz
channels, and one wide-band $800\,$MHz channel. 
The equivalent noise bandwidth of the wide-band channel after accounting for 
digitization losses and dropped end channels was $540\,$MHz.
Total on-source integration time for this field
was $43.1$ hours spread over 7 days during the period June 15 to
August 15, 1997 and  on Sept. 3, 1998.  
Using all the data to produce a naturally weighted image resulted 
in a beam-size of $34\2pr \times 26\2pr$ and a map {\it rms} of 
$91\,\mu{\rm Jy}\,{ \rm beam}^{-1}$ which corresponds to a 
Rayleigh-Jeans (RJ) temperature map {\it rms} of $160\,\mu$K.
We increased the brightness sensitivity of the image by applying a filter 
to the visibility ($u$-$v$) data.
Applying a Gaussian $u$-$v$ taper with a half-amplitude cutoff of $0.8\,\kl$ 
resulted in a beam-size of $101\2pr \times 97\2pr$ and an image {\it rms}
$185\,\mu{\rm Jy}\,{ \rm beam}^{-1}$ which corresponds to a RJ 
temperature map {\it rms} of $26\,\mu$K.
There appear to be several weak point sources in the images, but nothing 
resembling a CMB decrement. 

In the summer of 1998, we performed supplementary observations
with the OVRO millimeter array outfitted with the same $28.5\,$GHz receivers
used for the BIMA observations.
The large collecting area of the six $10.4\,$m OVRO telescopes and 
$2\,$GHz analog correlator provide excellent point source sensitivity. 
For a description of the OVRO system with Ka band HEMT
receivers see \markcite{Carlstrom}Carlstrom, Joy, \& Grego (1996).
In the week of June 15-22 1998, we performed three deep observations each 
centered on one of the three point sources reported in 
\markcite{Jones}Jones \ea (1997).
The coverage was not uniform, but for each of the three observations the array 
produced a naturally weighted beam $\sim 11\2pr \times 13\2pr$ and a map {\it rms} of
$\sim 100-150\,\mu{\rm Jy}\,{ \rm beam}^{-1}$.
Simultaneously fitting to all the OVRO and BIMA $u$-$v$, we detect three 
significant point sources in the field.  
The fluxes of the sources are all $\sim200\,\mu$Jy and their positions 
agree well with those of the three sources found in the Ryle observations. 
In \markcite{arcane}Holzapfel \ea (1999) we give a detailed description 
of the procedure for measuring the point sources and list the positions
and fluxes of the three significant sources.
The point source model is important for the anisotropy analysis presented 
in that work, but as we will show here, has little effect on the 
cluster emission constraints.

The combination of the OVRO and BIMA data do not reproduce the claimed decrement.
To quantify this result, we fit the $u$-$v$ data to the cluster emission model 
proposed by \markcite{Jones}Jones \ea (1997) to describe the Ryle data.
Fixing the position and shape parameters ($\beta=2/3$, $\theta_c=60\2pr$) of 
the spherical King model, we fit for the central decrement.
After subtracting the three detected point sources, we find a best fit central decrement 
$\Delta T_0=+49\pm85\,\mu$K.
This result is clearly inconsistent with the minimal model used to fit to 
the Ryle data of $\Delta T_0=-560\,\mu$K.

Because of the different spatial scales of the cluster and point sources,
the combination of the OVRO and BIMA data are able to simultaneously determine
the positions and fluxes of the point sources while constraining the cluster
emission model. 
We have repeated the fit for the central decrement while allowing 
the positions and fluxes of the three sources to simultaneously vary.
The point source fluxes and positions are virtually unchanged from the 
previously determined values and the central decrement is found to 
be $\Delta T_0=+48\pm90\,\mu$K.
In fact, removing the known point sources has no significant effect on the 
results; repeating the model fit without subtracting any point sources,
we find $\Delta T_0=+73\pm87\,\mu$K.
Therefore, the results we present are essentially independent of the point source 
model that we use.
Over the broad range of point source models we have considered, the BIMA 
and OVRO data are 
inconsistent with the cluster emission model of \markcite{Jones}Jones \ea (1997) 
at greater than $6\sigma$.

\begin{figure}[htb]
\plotone{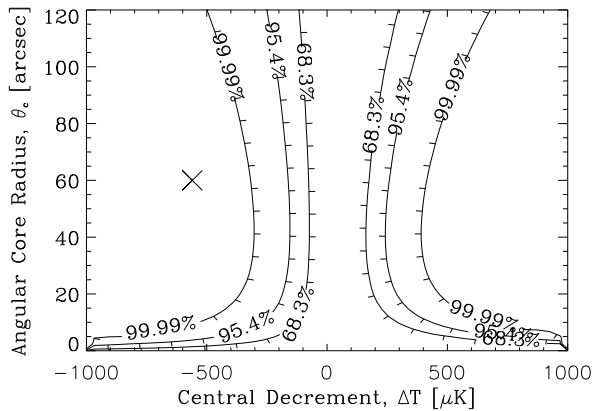}
\caption[]
{Confidence intervals for fits of the data from the BIMA and OVRO observations of 
PC1643+4631 to a spherical King model.
The `$\times$' marks the values of model parameters adopted in 
\markcite{Jones}Jones \ea (1997)
to describe their observed decrement. The BIMA data 
are inconsistent with this and any similar models at $ > 99.99\%$ confidence.}
\label{rylefig}
\end{figure}

To investigate what range of models for cluster emission the BIMA data are 
consistent with, we have fit the $u$-$v$ data to a grid of spherical 
King models (eqn.~\ref{eqn:sking}).
The values of $\beta$ and the angular core radius $\theta_c$ are largely 
degenerate, so we have fixed $\beta$ to a typical observed value of $2/3$.
Fitting the models to our data, we can generate confidence intervals in the two
free parameters, the central temperature decrement $\Delta T_0$ and the angular
core radius $\theta_c$.  In Figure~\ref{rylefig}, we show confidence intervals 
for a large range of model parameters.
For $\theta_c \sim 60\2pr$, only points near $\Delta T_0=0$ result in acceptable 
fits to the data.
The point marked with `$\times$' corresponds to the minimum King model used by 
\markcite{Jones} Jones \ea (1997) to describe the Ryle data: $\beta=2/3$,
$\theta_c = 60\2pr$, and $\Delta T_0 = -560\,\mu$K. 
This model is inconsistent with the BIMA data at grater than $99.99\%$ confidence.

Due to the East-West configuration of the Ryle array, the synthesized beam
is considerably elongated in declination.
Is it possible that weak cluster emission well matched to the Ryle beam could 
escape detection with BIMA?
To answer this question quantitatively, we have used the BIMA data to constrain highly 
elliptical cluster models which would be well matched to the Ryle beam. 
Fixing the axial ratio to 2.0 and the major axis core radius to $\theta_c=90\2pr$,
we find $\Delta T_0=+15\pm80\,\mu$K.
If we assume that $\Delta T_0=-560\,\mu$K, the BIMA data are inconsistent with 
this model for emission at greater than $7\sigma$.
We conclude that elongated emission can not account for the discrepancy between 
the Ryle and BIMA results.

\subsection{Discussion of PC1643+4631}
The x-ray results imply that any cluster capable of
producing the observed SZE decrement would have to be at a redshift of
$z>2.8$.
The optical observations are not as constraining, but show no evidence
of a massive cluster. 
The BIMA data we present are inconsistent with the model for 
the SZE emission proposed by \markcite{Jones}Jones \ea (1997) at 
$>99.99 \%$ confidence.
This result is insensitive to the details of the point source and
cluster emission models. 
Given the combination of the unlikely redshift range to which the x-ray 
constraints would push the cluster, and the significance of the BIMA 
non-detection, we find the evidence against a massive cluster in the direction 
of PC1643+4631A,B very convincing.

\section{VLA1312+4237}
\subsection{VLA Observations}
The Very Large Array (VLA) has been used to perform a 
sensitive search for CMB anisotropies on angular scales of $6\2pr$ to
$80\2pr$ (\cite{Partridge}). 
These observations consist of several hundred hours of 
observations at $8.44\,$GHz. 
The observed field, as defined by the primary beam half
power points, was $5.2\pr$ wide centered on the
coordinates $\alpha=13^{\rm h}\,12^{\rm m}\,17.4^{\rm s},\, 
\delta=+42^{\circ}\,38\pr\,15\2pr$ (J2000).  
This field was chosen to to be free of point sources with flux
density $S_{8.4\,\rm{GHz}} \ge 0.5\,$mJy. 
All positive features brighter than $7\,\mu$Jy ($4.7 \sigma$) were located in
the full resolution map and subtracted from the $u$-$v$ data.
The CLEANed $6\2pr$ resolution map had a {\it rms} noise of $1.5\,\mu$Jy,
making it the most sensitive radio map at any frequency or
resolution.
An isolated, negative flux feature 
approximately $30\2pr \times 65\2pr$ in size with a peak 
amplitude of $-250\,\mu$K was found in the residual map.
After point source subtraction, the significance of the feature was 
$5.5\, \sigma$ with approximately 680 independent beams in the Field
of View.
In \markcite{Richards}Richards \ea (1997), this feature is interpreted as the 
SZE signal of a distant cluster.
The image is extended in declination with considerable substructure. 
It has a northern decrement of $-13.9 \pm 3.3\,\mu$Jy ($-250 \pm 60\,\mu$K) in 
the $30\2pr$ resolution VLA image, and a southern decrement which is of 
similar amplitude.
The positions of these two features are given in Table~\ref{tab:sourcepos}.
\markcite{Richards}Richards \ea (1997) describe the observed signal with a spherical 
King model at the position 
of the Northern Peak with $\beta=2/3$, $\theta_c=15\2pr$, and central decrement 
$\Delta T_0=-250\,\mu$K.

\subsection{X-ray}
As related in \markcite{Richards}Richards \ea (1997), 
\markcite{Hu}Hu \& Cowie (1996) obtained a 
sensitive ROSAT HRI image of the region containing the source.
These observations were used to constrain the flux in the HRI
$0.1-2.4\,$keV band,
$f_x < 2 \times 10^{-14}\,{\rm erg\,cm}^{-2}\,{\rm s}^{-1}$
at the $3\sigma$ level.
Fixing $\beta=2/3$, Partidge \ea (1997) used the observed CMB 
decrement to estimate possible cluster parameters, specifically
the cluster x-ray luminosity $L_x \sim 2 \times 10^{44}\,{\rm erg s}^{-1}$
and gas mass $M_{gas} \sim 10^{13}\,M_{ \sun}$.
The lack of observed x-ray emission is used to argue that the
cluster, if it exists, must be at redshift $z > 1.5$.

\subsection {Optical} 
Optical images of the region containing the decrement had been 
produced as part of the HST Medium Deep Survey.
Two quasars at the identical redshift of $z=2.561$ separated by 
$100\2pr$ were discovered in the image.
For the two quasars to be the lensed images of a single object, 
an intervening mass of $\sim 10^{15} M_{\sun}$ would be required.
Also, the spectra of the two objects appear to be radically different.
The presence of these two quasars in such close proximity led 
\markcite{Richards}Richards \ea (1997) to propose that
they might be members of a cluster at that redshift.

\markcite{Campos} Campos \ea (1998) have searched the vicinity of the
quasar pair for evidence of galaxy clustering.
They find 56 Ly-$\alpha$ emitting candidates in a $8\pr \times 14\pr$ field.
Four of the five spectroscopically confirmed objects form a $3\pr$ 
($\sim 3\,$Mpc) filament with velocity dispersion $1000\,{\rm km\,s}^{-1}$.
This is several times the velocity dispersion one would expect for an 
unbound system of scale $\sim 3\,$Mpc.
However, the velocity dispersion of only 4 objects cannot be used to
identify a virialized system and many more spectra will be needed 
before this claim can be taken seriously.

\subsection{BIMA observations}
We observed VLA1312+4237  with our cm-wave receivers mounted on 9 of the
$6.1\,$m telescopes of the BIMA array. Total on-source integration time
was $35.5$ hours spread over 8 separate days during the period June 15 to
August 15, 1997.
The configuration of the array was identical to that used for the observations
of PC1643+4631.
Using all the data to produce a naturally weighted image resulted in a 
beam-size of $33\2pr \times 25\2pr$ and an image 
{\it rms} of $100\,\mu{\rm Jy}\,{ \rm beam}^{-1}$ which
corresponds to a RJ temperature map {\it rms} of $169\,\mu$K. 
Applying a Gaussian $u$-$v$ taper with a half-amplitude cutoff of $0.8\,\kl$ 
resulted in a beam-size $98\2pr \times 94\2pr$ and an image {\it rms}
$189\,\mu{\rm Jy}\,{ \rm beam}^{-1}$
which corresponds to a RJ temperature map {\it rms} of $29\,\mu$K.
We also created maps limited in $u$-$v$ range to search for point source 
emission.
There is no evidence for the presence of point source emission or a CMB 
decrement in any of the images. 

The $u$-$v$ coverage of the VLA and BIMA data sets is complementary with 
little overlap. 
The brightness sensitivity of the BIMA data is highest at angular scales of 
$50-120\2pr$ while the sensitivity of the VLA data is highest at scales of $6-30\2pr$.
We have investigated if the spherical King model suggested by Richards et al.\ (1997) 
is consistent with the BIMA data.  
Assuming the cluster to be described by model parameters $\beta =2/3$ and 
$\theta_c=15\2pr$, we find a best fit value for the central decrement of  
$\Delta T_\circ = +31\pm151\,\mu$K.
For larger core radii, the constraints are stronger.
Fixing $\theta_c=30\2pr$, we find $\Delta T_\circ = +38\pm118\,\mu$K.

\begin{figure}[htb]
\plotone{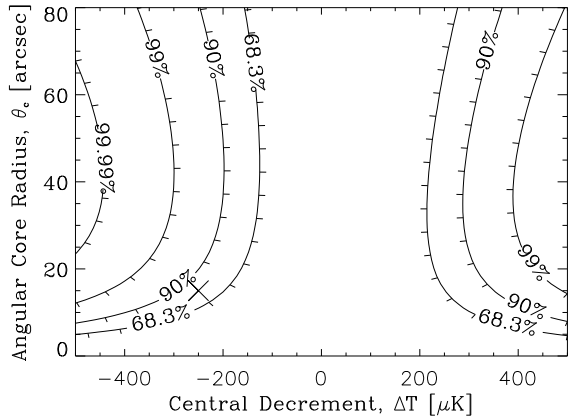}
\caption[]
{Confidence intervals for fits of the data from BIMA observations of VLA1312+4237 
to a spherical King model.
The `$\times$' marks the values of model parameters adopted in 
\markcite{Richards}Richards \ea (1997) to describe their observed decrement. 
The BIMA data are inconsistent with this model at $\sim 80\%$ confidence.}
\label{vlafig}
\end{figure}

In Figure~\ref{vlafig}, we show confidence intervals for fits to a range
of spherical King (eqn.~\ref{eqn:sking}) models for the cluster emission.
Again, we have fixed $\beta=2/3$.
The `$\times$' marks the parameters proposed by Richards \ea (1997),
which are $\Delta T_\circ = -250\,\mu$K, and $\theta_c=15\2pr$ with the model 
centered on the position of the ``Northern Peak''. 
The proposed cluster emission model is inconsistent with the BIMA data, although with
only $\sim 80\%$ confidence.

The feature observed with the VLA appears to be quite elongated  
in declination.
We have repeated the fits for cluster emission models which are better matched 
to the morphology of the reported VLA feature.
For these observations, we have fixed the source position midway between the
northern and southern peaks of the VLA image.
Assuming the cluster profile to be described by an elliptical King model
with an axial ratio of two and a major axis described by a core radius
of $\theta_c=30\2pr$ aligned in the North-South direction, we find 
$\Delta T_\circ = +146\pm125\,\mu$K.
Assuming the central decrement to be $\Delta T_0=-250\,\mu$K, the BIMA 
data are inconsistent with this elliptical model for the cluster emission 
at greater than $3\sigma$.

\subsection{Discussion of VLA1312+4237}
We find no evidence for a massive cluster in the 
VLA1312+4237 field, but cannot rule out its presence with high confidence.
The x-ray and optical data do not constrain the presence of the cluster at
$z=2.61$ as proposed by \markcite{Richards}Richards \ea (1997), and there
may be some evidence of structure in the distribution of background galaxies 
in addition to the two quasars in the vicinity of the proposed decrement.
The angular scales probed by the VLA and BIMA instruments overlap very little,
and comparison of the two data sets must be done through a model for the 
cluster emission. 
The model proposed by \markcite{Richards}Richards \ea (1997) for the cluster 
emission is inconsistent with the BIMA data at $\sim 80\%$ confidence.
Further observations with the OVRO array, or a more extended configuration 
of the BIMA array, would permit a more significant test of the
existence of a SZE decrement.

\section{Conclusions}
We have observed two fields containing claimed decrements in the CMB.
In both cases, we do not detect the reported decrements
and have been able to, with varying confidence, rule out emission described 
by the cluster models proposed by the authors of the detection papers.
The BIMA observations are inconsistent with the cluster model put forth by 
\markcite{Richards}Richards \ea (1997) for the decrement in the direction 
of VLA1312+4237, but only at $\sim 80\%$ confidence.

Due to the similarity of the angular scales to which the BIMA and Ryle 
telescopes are sensitive, the constraints are particularly strong in the case 
of PC1643+4631.
The BIMA data are found to be inconsistent with the proposed emission model at 
greater than $99.99\%$ confidence.
The x-ray and optical observations provide no support for a
massive cluster in the direction of PC1643+4631.
In a companion paper, we use the observations of both of these and five additional
fields to place limits on arcminute scale CMB anisotropies (\cite{arcane}).

We expect that the SZE will become a powerful tool for the
discovery of distant clusters.
The realization of this goal is likely to require not only an increase in 
sensitivity, but also a better understanding of the systematic errors 
associated with each experiment. 
Ultimately, multifrequency measurements with different instruments provide 
the strongest discrimination against foreground confusion and 
systematic errors.

We would like to thank the most excellent staff of the BIMA and OVRO 
observatories for their assistance with the observations.
Thanks to Cheryl Alexander for her help in the construction of
the cm-wave receivers and to Asantha Cooray and Sandy Patel for 
assistance with the OVRO and BIMA observations.
Radio Astronomy with the OVRO millimeter array is supported by NSF
grant AST 96-13717.
The BIMA millimeter array is supported by NSF grant AST 96-13998.
JEC acknowledges support from a NSF-YI grant and the David and Lucile
Packard Foundation.
EDR and LG acknowledge support from a NASA GSRP fellowship.
This work is supported in part by NASA LTSA grant number NAG5-7986.

\markright{REFERENCES}

\end{document}